\documentstyle[preprint,tighten,floats,prd,aps]{revtex}
\input epsf.tex
\def\DESepsf(#1 width #2){\epsfxsize=#2 \epsfbox{#1}}
\begin{document}
\preprint{\vbox{
\hbox{OITS 565}
\hbox{UW/PT 94-16}
\hbox{hep-ph/9412342}}  }
\draft

\title{Triply Differential Jet Cross Sections\\
for Hadron Collisions at Order $\alpha_s^3$ in QCD}
\author{Stephen D. Ellis}
\address{Department of Physics,\\
University of Washington, Seattle, WA 98195}
\author{Davison E.\ Soper}
\address{Institute of Theoretical Science,\\
University of Oregon, Eugene, OR 97403}
\date{21 December 1994}
\maketitle

\begin{abstract}
We discuss cross sections for $hadron+hadron \to 2\ jets + anything$ in
which three jet variables are measured. Such cross sections are useful
especially for determining parton distributions. We define a new cross
section $d\sigma/dX_A\,dX_B\,d\eta_*$ for which the perturbation
theory is nicely behaved even in the kinematic regime where the parton
distributions are probed at large momentum fractions. The cross section
$ d\sigma/dE_{T1}\, d\eta_1\, d\eta_2$, which has been used in the
past, is not so well behaved in this region. We calculate these cross
sections at order $\alpha_s^3$ in QCD.
\end{abstract}



\narrowtext

The experimental investigation of jet production in high energy hadron
collisions provides a direct view of the underlying process,
parton-parton scattering. It thus provides an opportunity to test
quantum chromodynamics (QCD) in some detail. The one jet inclusive
cross section $d\sigma/dE_T$, where $E_T$ is the transverse energy of
the jet, provides an excellent probe of any possible breakdown of QCD
at short distances \cite{onejet}. The two jet inclusive cross section
$d\sigma/dM_{JJ}$, where $M_{JJ}$ is the invariant mass of the two jet
system, plays a similar role, and also tests for possible resonances
that might be produced in parton-parton scattering \cite{mjj}. The two
jet angular distribution $[d\sigma/dM_{JJ}\, d\eta_*]
/[d\sigma/dM_{JJ}]$, where $\eta_*$ is half the rapidity difference
between the two jets, probes the angular dependence of the Feynman
diagrams for parton-parton scattering \cite{etastar}. These tests have
confirmed QCD quite convincingly  in a transverse momentum range from
30 GeV all the way to 400 GeV.

It is possible to study inclusive two jet production in even more
detail. In a Born level description, the physical process is the
scattering of two partons. Each of the two outgoing partons is
described by three variables. However, transverse momentum conservation
eliminates two of the six total variables, while symmetry about the
beam axis makes one variable superfluous. Thus there are three
independent variables. So at the Born level one can define, at most, a
triply differential cross section for two jet production. For instance,
the rapidities $\eta _1$ and $\eta _2$ of the two partons and the
transverse energy $E_{T1}$ of one of them can serve as the three
independent variables, leading to a cross section $d\sigma
/dE_{T1}\,d\eta _1\,d\eta _2$. This choice of the three variables,
however, is not unique and one may consider other sets of three
variables to describe the two outgoing partons and thus other triply
differential cross sections. One must also specify how the cross
section definition is extended from the case of two partons to the real
case of many particles. In this paper we discuss triply differential
two-jet cross sections with the goal of using such a cross section to
help determine parton distribution functions, particularly at large
momentum fractions. We attempt to define the cross section such that
the next-to-leading-order contributions are small compared to the
Born-level contributions over the entire allowed phase space.

We begin with the cross section $d\sigma /dE_{T1}\,d\eta _1\,d\eta _2$,
which has been measured by both the CDF \cite{CDFtriplydiff} and D0
\cite{D0triplydiff} groups at Fermilab and calculated at order
$\alpha_s^3$ by Giele, Glover, and Kosower \cite{Giele}. We define jets
according to the standard cone algorithm \cite{snowmass}, supplemented
by certain algorithms for dealing with overlapping jet cones
\cite{onejet,overlaps}. Each jet is labeled with variables $E_T$,
$\eta$ and $\phi$. Here $E_T$ is the sum of the absolute values of
the transverse momenta of all the particles in the jet cone. (CDF
adopts a slightly different definition of $E_T$ \cite{onejet}). The
variables $\eta $ and $\phi $ are the $E_T$-weighted averages of the
rapidities and azimuthal angles of all the particles or calorimeter
towers in the jet cone. In each event, we pick the two jets with the
highest transverse energies. One of the two jets is defined to be the
trigger jet, with transverse energy and rapidity $E_{T1},\,\eta _1$.
Then $\eta _2$ is the rapidity of the other jet. Since there are two
ways to choose which jet is the trigger jet, each event contributes to
two bins of the cross section.

\begin{figure}[htb]
\centerline{ \DESepsf(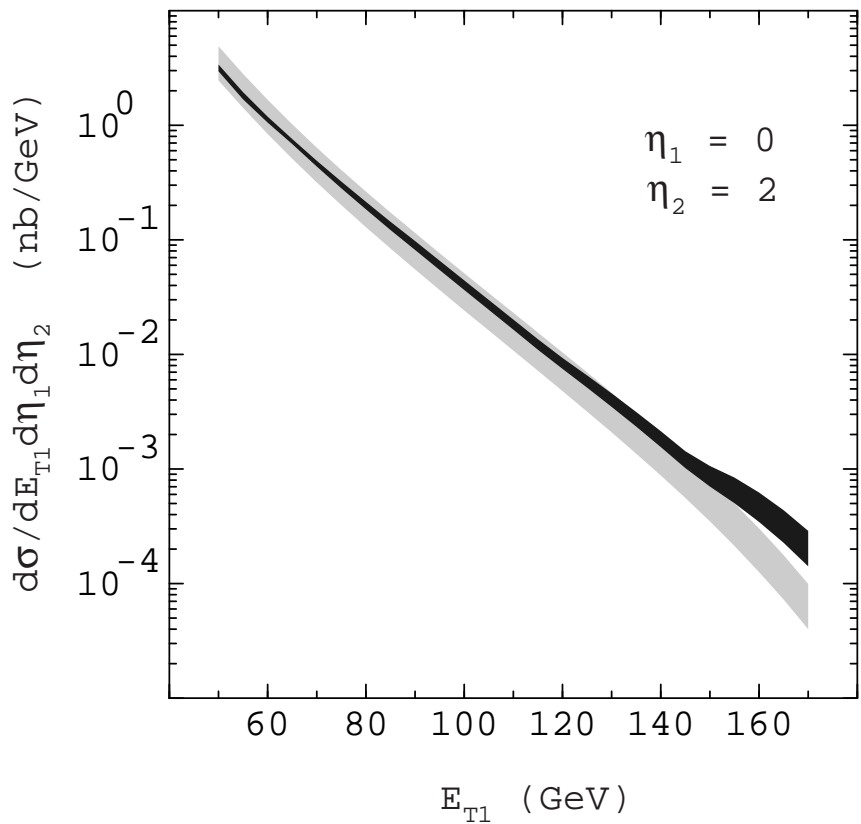 width 12 cm) }
\caption{ The cross section $d \sigma/dE_{T1}\,d\eta_1\,d\eta_2$ as
defined in the text plotted against $E_{T1}$ for $\eta_1 = 0$ and
$\eta_2 = 2$. Dark band: full order $\alpha_s^3$ cross section. Light
band: Born cross section (order $\alpha_s^2$). The bands reflect the
variation of the calculated cross section with the renormalization and
factorization scales. We use the CTEQ2 parton distributions
\protect{\cite{CTEQ}}.}
\label{ETy1y2:fig}
\end{figure}

The agreement between theory and experiment for $d\sigma /dE_{T1}\,d\eta
_1\,d\eta _2$ is satisfactory, given the experimental and theoretical
errors \cite{CDFtriplydiff,D0triplydiff,Giele}. However, as we shall
see, the cross section has large theoretical errors in certain regions
of the jet-variable space. In Fig.~\ref{ETy1y2:fig}, we show the cross
section at $\eta _1=0$ and $\eta _2=2$ plotted against $E_{T1}$. We
plot both the Born cross section and the full order $\alpha_s^3$ cross
section as bands whose widths reflect the theoretical error. The
calculated cross section depends on an $\overline{{\rm MS}}$
renormalization scale, $\mu _{UV}$, describing the regulation of
ultraviolet divergences, and an $\overline{{\rm MS}}$ factorization
scale, $\mu _{CO}$, describing the regulation of collinear divergences.
We define these parameters to be $\mu _{UV}=A_{UV}\,E_{T1}/2$,
$\mu _{CO}=A_{CO}\,E_{T1}/2$. We choose $(A_{UV},A_{CO})=(1,1)$ as the
standard value. Then we determine the bands by calculating the cross
sections for $(\log (A_{UV}),\log (A_{CO}))$ chosen at eight points
around the circle $\log ^2(A_{UV})+\log ^2(A_{CO})=2\log ^2(2)=0.96$.
The upper and lower edges of the error bands correspond to the maximum
and minimum, respectively, of the cross section calculated in this
range of $ (A_{UV},A_{CO})$ values. Such an error band for the order
$\alpha_s^N$ cross section constitutes an estimate of the error
induced by omitting order $\alpha_s^{N+1}$ and higher contributions to
the cross section.

The results shown in Fig.~\ref{ETy1y2:fig} indicate that the theory is
well behaved in the central region, $50\ {\rm GeV}< E_{T1} < 140\ {\rm
GeV}$. However, the theory is {\em not} well behaved in the region
$140\ {\rm GeV}< E_{T1}$. The $\alpha_s^3$ corrections to the Born
cross section are large and so are the estimated errors. The reason is
simple \cite{Giele}. The allowed kinematic region for the Born cross
section for $\eta_1 = 0$ and $\eta_2 = 2$ is $E_{T1}< 215\ {\rm GeV}$.
Beyond this value of $E_{T1}$, the momentum fraction of the incoming
quark from hadron $A$ becomes larger than 1. Although the Born cross
section is zero for $E_{T1}> 215\ {\rm GeV}$, the full $\alpha_s^3$
cross section is not, as we will discuss below. Thus for
$E_{T1}> 215\ {\rm GeV}$, and even for some interval of $E_{T1}$ below
this value, the order $\alpha_s^3$ cross section is effectively a
leading order prediction. Generally speaking, leading order cross
sections have large estimated errors. This case is no exception. The
estimated error is greater than $\pm 15\%$ for $E_{T1} > 140\ {\rm
GeV}$. At the edge of our plot, the estimated error has reached $\pm
35\%$ and is still growing with $E_{T1}$.

The following general, if somewhat abstract, formalism will allow us to
understand how the allowed region at order $\alpha_s^N$ is related to
the definition of the jet cross section. In an order $\alpha_s^N$
calculation, one can have up to $N$ partons in the final state. These
partons can be described by $3N-3$ parameters $v_i$ in a space ${\cal
V}_N$. The definition of a specific triply differential cross section,
{\it i.e}. of a specific set of three jet-variables,  may be understood
as a set of maps of the $N$-parton space ${\cal V}_N$ into a three
dimensional space ${\cal V}_0$ of measured jet-variables. Equivalently,
we can view the definition as mapping the ${\cal V}_N$ into ${\cal
V}_2$, followed by a map of ${\cal V}_2$ into ${\cal V}_0$. Thus a point
$v$ in ${\cal V}_N$ is mapped into a point ${\cal M}_N(v) $ in ${\cal
V}_2$. Now there are many possible sets of maps ${\cal M}$ corresponding
to different choices of the three variables. The only restriction comes
from the requirement of infrared safety. When, in an $N$-parton
configuration, one parton becomes soft or collinear to one of the beams
or two partons become collinear, then that configuration must map to
the same point in ${\cal V}_2$ as the physically equivalent
($N-1$)-parton configuration. This will ensure the cancellation of the
relevant singularities in the perturbation theory.

Consider, now, the physically allowed region ${\cal A}_N$ in ${\cal
V}_N$. This is the region determined by $x_A<1$ and $x_B<1$, where the
incoming parton momentum fractions are
\begin{equation}
x_A=\sum_n{\frac{E_{Tn}}{\sqrt{s}}}\,e^{\eta _n},\quad
x_B=\sum_n{\frac{E_{Tn}}{\sqrt{s}}}\,e^{-\eta _n}.  \label{xdef}
\end{equation}
This region is mapped into a region ${\cal M}_N({\cal A}_N)$ in ${\cal
V}_2$ . How does ${\cal M}_N({\cal A}_N)$ compare to ${\cal A}_2$? The
infrared safety condition implies that
\begin{equation}
{\cal M}_N({\cal A}_N)\supseteq {\cal A}_2.  \label{contained}
\end{equation}
To see this, let $\tilde v$ be a two parton point in the allowed region
${\cal A}_2$. Then consider the physically equivalent point $v$ in
${\cal V}_N $ obtained by adding $N-2$ zero momentum partons to
$\tilde v.$ The infrared safety of the map implies that ${\cal
M}_N(v)=\tilde v$. Since $v\in {\cal A}_N,$ Eq.~(\ref{contained})
follows.

While Eq.~(\ref{contained}) holds for {\it any} infrared safe
definition of a triply differential cross section, the answer to the
question of whether ${\cal M}_N({\cal A}_N)$ is equal to ${\cal A}_2$ or
bigger than ${\cal A}_2$ depends on which definition one chooses.
Consider the particular cross section $d\sigma /dE_{T1}\,d\eta
_1\,d\eta _2$ defined earlier, and denote the corresponding maps as
${\cal M}_N^{{\rm I}}$. It is easy to see that the region ${\cal
M}_N^{{\rm I}}({\cal A}_N)$ is bigger than ${\cal A}_2$. For instance,
the three parton final state with
\begin{eqnarray}
E_{T1} &=&220\ {\rm GeV},\ \eta _1=0,\ \phi _1=0,
  \nonumber \\
E_{T2} &=&120\ {\rm GeV},\ \eta _2=2,\ \phi _2=\pi,
\label{numerical} \\
E_{T3} &=&100\ {\rm GeV},\ \eta _3=0,\ \phi _3=\pi  \nonumber
\end{eqnarray}
has $x_A=0.67$, $x_B=0.19$. This kinematically allowed state is mapped to
the two parton final state with
\begin{eqnarray}
E_{T1} &=&220\ {\rm GeV},\ \eta _1=0,\ \phi _1=0,
\nonumber \\
E_{T2} &=&220\ {\rm GeV},\ \eta _2=2,\ \phi _2=\pi .
\end{eqnarray}
This state has $x_B=0.14$ but $x_A=1.03$, so it is not in the
kinematically allowed region ${\cal A}_2$ and ${\cal M}_N^{{\rm
I}}({\cal A}_N)\supset  {\cal A}_2$.

The results depicted in Fig.~\ref{ETy1y2:fig} suggest that the cross
section $d\sigma /dE_{T1}\,d\eta _1\,d\eta _2$ provides a useful tool
for exploring QCD and testing parton distributions, but that there are
difficulties resulting from the fact that the allowed regions ${\cal
M}_N({\cal A}_N)$ at order $\alpha_s^N$ are larger than ${\cal A}_2$,
the region allowed at the Born level. Now the question arises, can we
define a triply differential jet cross section such that the maps of
all of the allowed regions ${\cal M}_N({\cal A}_N)$ equal ${\cal
A}_2$?  In this case the perturbative result can be well behaved over
the entire allowed region. The answer is that there are many such
solutions and one particular choice stands out as being particularly
simple.

We define a cross section $d\sigma /dX_A\,dX_B\,d\eta _{*}$ as follows.
Define jets according to the standard cone algorithm \cite{snowmass}. In
each event, pick the two jets with the highest transverse energies
$E_T$. Let
\begin{equation}
\eta _{*}=|\eta _1-\eta _2|/2,
\end{equation}
where $\eta _1$ and $\eta _2$ are the rapidities of these two leading
jets. (That is, $\eta $ is the $E_T$-weighted average of the rapidities
of all the particles or calorimeter towers in the jet
cone \cite{snowmass}.) Let
\begin{equation}
X_A=\sum_{i\in {\rm jets}}{\frac{E_{Ti}}{\sqrt{s}}}\ e^{\eta _i},\quad
X_B=\sum_{i\in {\rm jets}}{\frac{E_{Ti}}{\sqrt{s}}}\ e^{-\eta _i}.
\label{Xdef}
\end{equation}
The sum here runs over all of the particles (or calorimeter towers) in
either of the jet cones.

Let us call the maps corresponding to this definition ${\cal M}_N^{{\rm
II}}$. It is easy to see that the allowed regions ${\cal M}_N^{{\rm
II}}({\cal A}_N)$ are all the same. For an allowed $N$-parton
configuration, the true momentum fractions $x_A$ and $x_B$, as given in
Eq.~(\ref{xdef}), are less than 1. From Eq.~(\ref{Xdef}) we obtain
\begin{equation}
X_A\leq x_A, \quad X_B\leq x_B.
\end{equation}
Thus an allowed $N$-parton state must have $X_A<1$ and $X_B<1$. In the
two parton final state with the same jet variables, the momentum
fractions of the final state partons are precisely $X_A$ and $X_B$.
Thus an allowed $N$-parton final state is mapped into an allowed
two-parton final state. That is, ${\cal M}_N^{{\rm II}}({\cal
A}_N)\subseteq {\cal A}_2$. Since we already know from
Eq.~(\ref{contained}) that ${\cal M}_N^{{\rm II}}({\cal A}_N)\supseteq
{\cal A}_2$, we have
\begin{equation}
{\cal M}_N^{{\rm II}}({\cal A}_N)={\cal A}_2.
\end{equation}

The numerical example used earlier illustrates the point. We considered
the allowed three parton configuration with parameters $v$ given by
Eq.~(\ref {numerical}). With the new definition, this state has jet
variables $X_A=0.61$, $X_B=0.13$, $\eta _{*}=1$. The corresponding two
parton state $\tilde v={\cal M}_3^{{\rm II}}(v)$ is
\begin{eqnarray}
E_{T1} &=&166\ {\rm GeV},\ \eta _1=-0.23,\ \phi _1=0,
\nonumber \\
E_{T2} &=&166\ {\rm GeV},\ \eta _2=1.77,\ \ \ \, \phi _2=\pi
\end{eqnarray}
with $x_A=X_A=0.61$ and $x_B=X_B=0.13$. Thus with the new definition the
two parton state is kinematically allowed.

The variables $X_A$ and $X_B$ are conceptually straightforward. To
calculate $X_A$, for instance, we sum of the plus-components $k_i^+ =
(k_i^0 + k_i^3)/\sqrt2$ of the four-momenta of the particles inside
both jet cones and divide by the plus-component of hadron $A$'s
momentum. On the experimental side, we note that, for large rapidity
$\eta_i$, $E_{Ti}\exp(\eta_i)$ is approximately twice the energy of
particle $i$. Since calorimeters directly measure energy, the spreading
of a large rapidity jet over a region of rapidity in the detector
should not much affect the measurement of $X_A$.

The cross section $d\sigma /dX_A\,dX_B\,d\eta _{*}$ takes a very simple
form at the Born level:
\begin{eqnarray}
\lefteqn{X_A^2 X_B^2 s\ \frac{\,d\sigma }{dX_A\,dX_B\,d\eta _{*}}=} \\
&&\sum_{a,b}X_A f_{a/A}(X_A,\mu _A)\,X_B f_{b/B}(X_B,\mu _B)\,
\alpha_s^2(\mu _{UV})\,H_{ab}(\eta _{*}).
\nonumber
\end{eqnarray}
Here $f_{a/A}(X_A,\mu _A)$ is the parton distribution function for
finding a parton of type $a$ in hadron $A$, evaluated at some scale
$\mu _A$; $f_{b/B}(X_B,\mu _B)$ is the corresponding function for hadron
$B$. There are two powers of $\alpha_s$ evaluated at a scale $\mu
_{UV}$. Finally there is a function $H_{ab}(\eta _{*})$ that gives the
angular distribution for 2 parton $\to $ 2 parton scattering at c.m.\
energy squared $\hat s$:
\begin{equation}
{\frac{d\sigma (a+b\to a^{\prime }+b^{\prime })}{d\eta _{*}}}
={\frac{\alpha_s^2}{\hat s}}\ H_{ab}(\eta _{*}).
\end{equation}
The incoming partons are of type $a,b$ here and we sum over types of
outgoing partons. We see that the variables $X_A$, $X_B$, and
$\eta_{*}$ are well adapted to probing the factors in the Born-level
cross section, especially the parton distribution functions.

What of the scales $\mu$? Some sensible choices are suggested by the
kinematics of parton-parton scattering at the Born level, as seen in the
parton-parton c.m.\ frame. One choice might be half the energy of each
of the outgoing partons, $E/2 = [X_A X_B s]^{1/2}/4$, while another
might be half the transverse momentum of each of the outgoing partons,
$P_T/2 = [X_A X_B s]^{1/2}/ [4\cosh(\eta_*)]$. In our next-to-leading
order calculation, we use a compromise between these two approaches:
\begin{eqnarray}
\mu_{UV}&=&A_{UV} {\ \sqrt{X_A X_B s} / [4 \cosh (0.7\eta_*)]},
\nonumber \\
\mu_{A}&=&A_{CO} {\ \sqrt{X_A X_B s} / [4 \cosh ([1-X_A]\eta_*)]},
\nonumber \\
\mu_{B}&=&A_{CO} {\ \sqrt{X_A X_B s} / [4 \cosh ([1-X_B]\eta_*)]}.
\end{eqnarray}
Here the factors $A_{UV}$ and $A_{CO}$ are adjustable. Any choices of
these $A$'s that are of order 1 would be reasonable.

\begin{figure}[htb]
\centerline{ \DESepsf(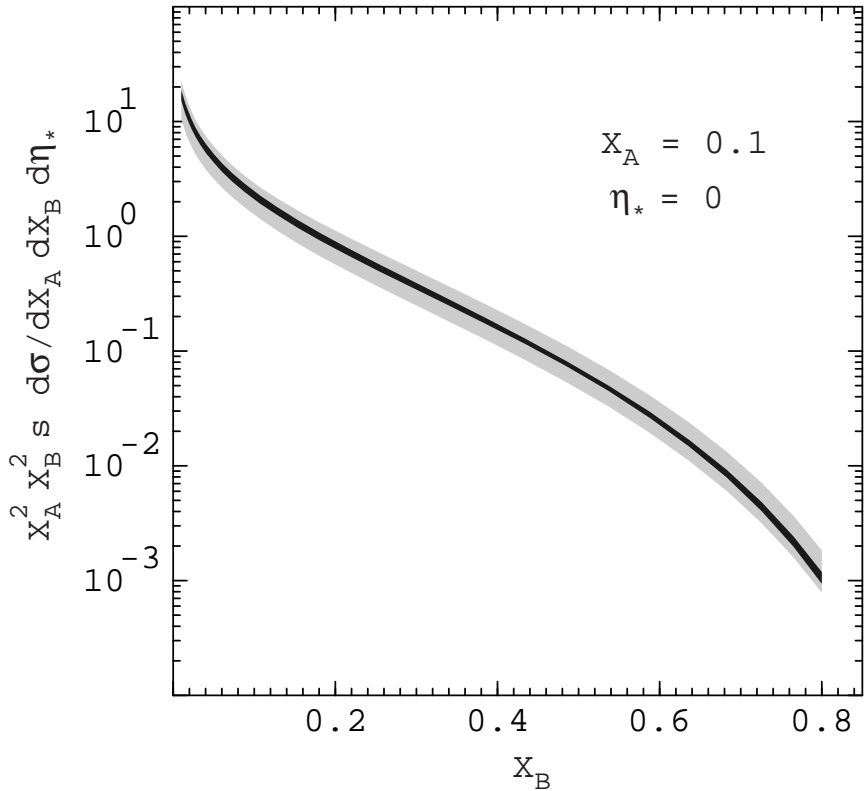 width 12 cm) }
\caption{ The cross section $d\sigma/dX_A\,dX_B\,d\eta_*$ as defined in
the text plotted against $X_B$ for $X_A = 0.1$ and $\eta_* = 0$. Dark
band: full order $\alpha_s^3$ cross section. Light band: Born cross
section (order $\alpha_s^2$). The bands reflect the variation of the
calculated cross section with the renormalization and factorization
scales. We use the CTEQ2 parton distributions \protect{\cite{CTEQ}}.}
\label{XAXBetastar:fig}
\end{figure}

In Fig.~\ref{XAXBetastar:fig}, we show the cross section
$d\sigma/dX_A\,dX_B\,d\eta_*$ at $X_A = 0.1$ and $\eta_* = 0$ plotted
against $X_B$. We plot both the Born cross section and the full order
$\alpha_s^3$ cross section, in each case as a band whose width reflects
the theoretical error as determined by choosing
$(\log(A_{UV}),\log(A_{CO}))$ at eight points around the circle
$\log^2(A_{UV}) + \log^2(A_{CO}) = 2\log^2(2) = 0.96 $, just as in
Fig.~\ref{ETy1y2:fig}.

The results shown in Fig.~\ref{XAXBetastar:fig} indicate that the
theory is well behaved in the entire $X_B$ range shown, even out to
quite large $X_B$. The $\alpha_s^3$ corrections are not large and the
$\alpha_s^3$ calculation has small estimated errors from higher order
contributions, about 10\%.

Our calculations in this paper are based on the subtraction algorithm
and corresponding computer code for calculating next-to-leading order
jet cross sections described in Ref.\ \cite{algorithm}. The desired
definitions of jets and jet variables are inserted into certain small
subroutines in the code. A new feature compared to previous versions of
the code is that the program directly calculates the so-called
$K$-factor, the ratio of the full order $\alpha_s^3$ cross section to
the Born cross section. This has technical advantages stemming from the
fact that the $K$-factor is normally nearly constant as a function of
the three jet variables, at least for an appropriately defined cross
section as here.

We conclude with a caveat. We have seen that the cross section
$d\sigma/dX_A\,dX_B\,d\eta_*$ is reasonably well behaved for small
$1-X_A$ or $1-X_B$. Nevertheless, the example of the Drell-Yan cross
section teaches us that the perturbation series will contain terms with
factors of $\log(1-X_A)$ and $\log(1-X_B)$. A summation of these
logarithmic terms will be useful, and perhaps necessary, in order to
explore with precision the small $1-X_A$ and $1-X_B$ regions.
Substantial progress along these lines has been made in the Drell-Yan
case \cite{Sterman}.

This work was supported by the United States Department of Energy. We
thank E.\ Kovacs, S.\ Kuhlmann, H.\ Weerts and Z.\ Kunszt for helpful
conversations. We thank W. Giele for help in comparing our results to
those of Ref.~\cite{Giele}.

\end{document}